\documentclass[journal]{IEEEtran}
\usepackage{amsmath,amssymb,amsfonts}
\usepackage{graphicx}
\usepackage{multirow}
\usepackage{cite}
\usepackage{booktabs}
\usepackage{siunitx}
\usepackage{url}
\usepackage{bbm} % or \usepackage{dsfont}
\begin{document}

\title{Efficient Coherence Inference Using the Demodulated Band Transform and a Generalized Linear Model}

\author{Md~Rakibul~Mowla, 
        Sukhbinder~Kumar, 
        Ariane~E.~Rhone, 
        Brian~J.~Dlouhy,  
        and Christopher~K.~Kovach%
%\thanks{Manuscript submitted to \emph{IEEE XXXX} on MONTH DD, YYYY.}%
\thanks{This work was supported in part by the National Institutes of Health (NIH), National Institute of Neurological Disorders and Stroke under Grant 5K08NS112573-02, and by the National Institute on Deafness and Other Communication Disorders under Grant 5R01DC004290.}%
\thanks{M.R.~Mowla, S.~Kumar, and A.E.~Rhone are with the Department of Neurosurgery, University of Iowa, Iowa City, IA 52242, USA (e-mail: mdrakibul-mowla@uiowa.edu).}%
%\thanks{M.A.~Howard~III is with the Departments of Neurosurgery and Neurology, and with the Iowa Neuroscience Institute, University of Iowa, Iowa City, IA 52242, USA.}%
\thanks{B.J.~Dlouhy is with the Departments of Neurosurgery and Pediatrics, University of Iowa, Iowa City, IA 52242, USA.}%
\thanks{C.K.~Kovach is with the Department of Neurosurgery, University of Nebraska Medical Center, Omaha, NE 68198, USA(e-mail: ckovach@unmc.edu).}%
}

\maketitle

\begin{abstract}
Statistical significance testing of neural coherence is essential for distinguishing genuine cross-signal coupling from spurious correlations. A widely accepted approach uses surrogate-based inference, where null distributions are generated via time-shift or phase-randomization procedures. While effective, these methods are computationally expensive and yield discrete $p$-values that can be unstable near decision thresholds, limiting scalability to large EEG/iEEG datasets. We introduce and validate a parametric alternative based on a generalized linear model (GLM) applied to complex-valued time--frequency coefficients (e.g., from DBT or STFT), using a likelihood-ratio test. Using real respiration belt traces as a driver and simulated neural signals contaminated with broadband Gaussian noise, we perform dense sweeps of ground-truth coherence and compare GLM-based inference against time-shift/phase-randomized surrogate testing under matched conditions. GLM achieved comparable or superior sensitivity while producing continuous, stable $p$-values and a substantial computational advantage. At $80\%$ detection power, GLM detects at $C \approx 0.25$, whereas surrogate testing requires $C \approx 0.49$, corresponding to an approximately $6$--$7$\,dB SNR improvement. Runtime benchmarking showed GLM to be nearly $200\times$ faster than surrogate approaches. These results establish GLM-based inference on complex time--frequency coefficients as a robust, scalable alternative to surrogate testing, enabling efficient analysis of large EEG/iEEG datasets across channels, frequencies, and participants.
\end{abstract}

% (Optional, IEEE Access)
\begin{IEEEkeywords}
Neural coherence, surrogate testing, demodulated band transform (DBT), generalized linear model (GLM), circular time shift, phase randomization.
\end{IEEEkeywords}

\section{Introduction}

\subsection{Problem Context: Neural Synchrony \& Coherence}
Neural oscillations play a central role in coordinating activity across distributed brain regions \cite{mackay1997synchronized}. A widely used measure for quantifying these interactions is the \textit{spectral coherence function}, which assesses the strength of frequency-specific coupling between two signals \cite{LaRocca2014}. Coherence has been extensively applied in studies using scalp electroencephalography (EEG), intracranial EEG (iEEG), and local field potentials (LFP) to investigate functional connectivity, identify biomarkers, and explore brain–body integration \cite{lachaux1999measuring,LaRocca2014,Busonera2018}. High coherence at a specific frequency often reflects shared information processing or common rhythmic modulation, making it particularly valuable for studying driver–channel relationships such as respiration-locked neural coupling \cite{Dias2025, Nakamura2024, herrero2018breathing}. However, coherence values can be inflated by shared noise, volume conduction, and narrow-band oscillations unrelated to genuine synchronization. Statistical testing is therefore essential to separate true interactions from spurious correlations.

\subsection{Current Practice: Surrogate-Based Significance Testing}
The standard approach for assessing coherence significance relies on \textit{surrogate data analysis} \cite{schreiber2000surrogate,Faes2004}. Among spectrum-preserving methods, Fourier-transform (\emph{phase-randomized}) surrogates are widely recommended because they preserve the amplitude spectrum while destroying phase relations, yielding frequency-dependent thresholds that adapt to local spectral structure \cite{schreiber1996improved,Faes2004}. In this work, we benchmark two spectrum-preserving baselines under matched conditions: (i) \emph{circular time shifts} (lag surrogates) applied to analytic narrowband (DBT) coefficients, and (ii) \emph{phase-randomized} (FT) surrogates. Other surrogate families exist—IID/random-shuffle \cite{palus1997detecting}, autoregressive (AR) \cite{prichard1994generating}, and amplitude-adjusted AAFT/IAAFT \cite{theiler1992testing,schreiber1996improved}—but are not evaluated here (see \cite{Faes2004} for comparisons). Despite their robustness, surrogate procedures are computationally intensive (hundreds to thousands of realizations per channel–frequency) and yield discrete $p$-values bounded by $1/(n_{\mathrm{perm}}{+}1)$, motivating a scalable parametric alternative.

\subsection{From Surrogates to a Parametric Framework}
To address these limitations, we represent signals with a complex time–frequency decomposition that yields narrowband analytic coefficients (e.g., the demodulated band transform, DBT \cite{kovach2016}, though the approach applies equally to STFT and related transforms). On these coefficients, we propose a generalized linear model (GLM) framework for testing coherence per frequency band: we fit a GLM to the complex-valued coefficients and perform a likelihood–ratio test with $\mathrm{df}=2$ (reflecting the real and imaginary degrees of freedom). As emphasized by \cite{Faes2004}, spectrum-preserving surrogates (FT/AR) are preferred for coherence because they yield frequency-dependent thresholds that track local spectral structure; our approach attains comparable benefits \emph{without} resampling by operating parametrically on DBT coefficients. Inspired by prior parametric work on oscillatory coupling \cite{vanwijk2015}—which used GLMs for phase–amplitude coupling (PAC)—we extend this framework to \emph{coherence}: frequency-specific linear dependence between a driver and a neural channel, tested directly on complex DBT coefficients. To our knowledge, this is the first work to propose a parametric framework for coherence testing and to systematically validate GLM-based inference against spectrum-preserving surrogate methods under matched physiological drivers and simulation settings.

The aforementioned prior work resorts to surrogate testing on coherence values in part because coherence, as the magnitude of an asymptotically normal value (coherency), does not have an asymptotically normal distribution under the null hypothesis.
Although parametric approaches may also assume the correct asymptotic distribution for the  magnitude of a complex Gaussian variable \cite{bartz2020beyond},  we propose a simpler solution motivated by the observation that coherence is identical in form to Pearson's correlation coefficient: like the conventional correlation coefficient, it is a measure of the strength of the linear relationship between two signals at a defined frequency. The appropriate inferential test to accompany such a measure is on the \emph{presence} of a linear relationship, as handled by standard procedures of linear modeling. The conventional generalized linear model extends easily to the case when dependent and/or independent variables are complex-valued \cite{wooding1956multivariate, van2002multivariate}, which are in both cases isomorphic to bivariate real variables in real and imaginary parts. We therefore propose that the underlying complex (or equivalent bivariate) linear model is the most natural basis for inferential testing on coherence.

\subsection{Overview of Contributions}
Although DBT provides a stable spectral representation and GLM offers a principled alternative to surrogate-based testing, there has been no systematic validation for driver–channel coherence under matched conditions. We close this gap using controlled simulations driven by real respiration traces and noise-controlled observations to quantify sensitivity, agreement with surrogate tests, and computational efficiency.

We simulate observations in which a respiration-driven component is embedded in broadband Gaussian noise, scaling the noise to span a dense range of \emph{true} coherence values. Coherence is computed from complex band-limited coefficients (DBT with bandwidth $\mathrm{BW}=0.2$\,Hz over a record of length $T\approx 367$\,s, yielding a time–bandwidth product $\mathrm{TBP}=\mathrm{BW}\cdot T\approx 73.4$). Significance at the pre-specified breathing bin is assessed using (i) a GLM likelihood-ratio test ($\mathrm{df}=2$) and (ii) two spectrum-preserving surrogate baselines: circular time shifts and phase randomization. Performance is summarized by detection power, ROC analysis with a fixed control frequency, and runtime scaling.  

In terms of sensitivity, GLM and phase randomization achieved nearly identical detection power and ROC performance (AUC $=0.927$ for both), both outperforming circular shift (AUC $=0.848$). The critical advantage of GLM lies in runtime: at $n_{\mathrm{perm}}{=}2000$, GLM completed in $0.041$\,s compared to $8.022$\,s for circular shift and $7.558$\,s for phase randomization—a speedup of $\sim$$198\!\times$ and $186.5\!\times$, respectively. GLM also avoids the $p$-value floor imposed by finite surrogate counts (e.g., $1/(n_{\mathrm{perm}}{+}1)$). As expected, absolute thresholds shift with the effective sample size set by the time–bandwidth product.

These results demonstrate that GLM-based inference on complex time–frequency coefficients is a statistically valid and computationally efficient alternative to surrogate testing. While prior work has applied parametric modeling in related contexts, coherence significance testing has remained largely reliant on surrogate procedures. Here we provide, to our knowledge, the first systematic validation of a GLM-based framework against spectrum-preserving surrogates under matched physiological drivers and simulation settings. Sensitivity is on par with phase randomization, but GLM offers substantial computational savings and continuous $p$-values, making it particularly suitable for large EEG/iEEG datasets. Section~\ref{sec:methods} details the observation model, DBT-based coherence, and the three tests; Section~\ref{sec:results} presents the simulation outcomes; Section~\ref{sec:discussion} discusses implications and limitations; and Section~\ref{sec:conclusion} concludes the paper.

%=============================================================================
\section{Methods} \label{sec:methods}
\subsection{Observation Model and Target Coherence}
A respiration belt trace $x(t)$, sampled at $F_s=\SI{250}{Hz}$, served as the driver. We applied the demodulated band transform (DBT) \cite{kovach2016} with bandwidth $B=\SI{0.2}{Hz}$ and a frequency upsampling factor $U_{\mathrm{fx}}$ (here $U_{\mathrm{fx}}=3$). The peak breathing frequency $f_{\mathrm{br}}$ was determined from the DBT of the driver.

To construct observations with prescribed coherence levels, we generated $y(t)=x(t)+n(t)$, where $n(t)$ is Gaussian white noise scaled so that the coherence at $f_{\mathrm{br}}$ equals a target $C_{\text{true}}$. Writing $P_x$ for the (DBT) driver power at $f_{\mathrm{br}}$, the required noise standard deviation is
\[
\sigma_n(C_{\text{true}}) \;=\; \sqrt{\Big(\tfrac{1}{C_{\text{true}}^2}-1\Big)\,P_x \,\frac{F_{s}}{2B}\,\frac{(U_{\mathrm{fx}}+1)}{N}},
\]
where $N$ is the length of the driver sequence after resampling. Sweeping $C_{\text{true}}\in[0.001,0.999]$ (step $10^{-4}$), we added independent noise realizations at each target, applied DBT to obtain $A_Y(t,f)$, and analyzed the resulting $A_X,A_Y$ arrays ($T\times F\times N_{\text{coh}}$). The surrogate resolution floor was $p_{\min}=1/(n_{\mathrm{perm}}+1)$, with $n_{\mathrm{perm}}=2000$ permutations used to generate surrogate null distributions.

\subsection{DBT-based Coherence Estimation}
\begin{figure}[t]
  \centering
  \includegraphics[width=0.95\linewidth]{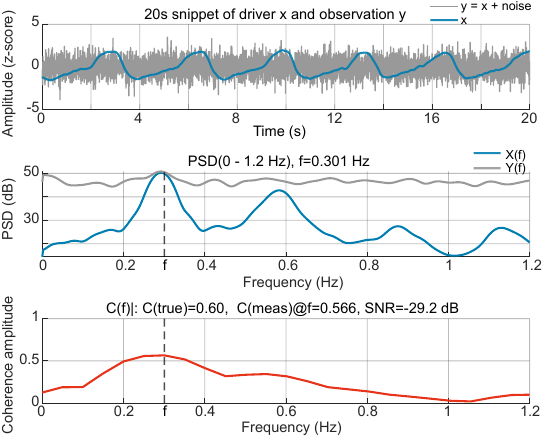}
  \caption{Representative driver and observation. \textbf{Top:} 20\,s snippet of the respiration driver $x(t)$ (blue, z-scored) and simulated observation $y(t)=x(t)+n(t)$ (gray). \textbf{Middle:} Power spectral densities of $x$ and $y$ (0--1.2\,Hz, dB scale; Welch); the dashed line marks the breathing frequency $f_{\mathrm{br}}$. \textbf{Bottom:} Coherence amplitude $C(f)$ computed from DBT coefficients via~\eqref{eq:coh}. Example shown at $C_{\text{true}}=0.60$; the DBT-measured observed coherence at $f{=}0.3$\,Hz is $C_{\text{meas}}=0.566$. Broadband SNR $= -29.2$\,dB.}
  \label{fig:samp}
\end{figure}

We estimate coherence using the demodulated band transform (DBT) \cite{kovach2016}. Signals are decomposed into overlapping complex-valued bands spanning 0–1.2~Hz with a 0.2~Hz bandwidth. For each band $f$, let $\{X_t(f)\}$ and $\{Y_t(f)\}$ denote the complex DBT coefficients of the driver and observation over $t=1,\ldots,T$ DBT samples (${}^*$ denotes complex conjugation). The classical coherence amplitude is
\begin{equation}
C(f) \;=\; \frac{\big|\sum_{t=1}^T X_t(f)\,Y_t^{*}(f)\big|}
{\max\!\left(\sqrt{\Big(\sum_{t=1}^T |X_t(f)|^2\Big)\Big(\sum_{t=1}^T |Y_t(f)|^2\Big)},\,\varepsilon\right)} ,
\label{eq:coh}
\end{equation}
where $\varepsilon$ is a small constant for numerical stability (set to MATLAB’s \texttt{eps} in our implementation). Unless stated otherwise, summaries use the bin nearest the breathing frequency $f_{\mathrm{br}}$. An illustrative example of the driver/observation, their PSDs, and the DBT-based coherence $C(f)$ computed via~\eqref{eq:coh} is shown in Fig.~\ref{fig:samp}.

\subsection{Significance Testing}
\textbf{GLM (parametric).}
For each band $f$, stack the complex DBT samples into $\mathbf{y}(f)\in\mathbb{C}^{T}$ (from $Y$) and $\mathbf{x}(f)\in\mathbb{C}^{T}$ (from $X$), and fit
\[
\mathbf{y}(f) \;=\; \beta(f)\,\mathbf{x}(f) \;+\; \varepsilon(f), \qquad \varepsilon(f)\sim\mathcal{CN}(0,\Sigma_f).
\]
Let $r_0(f)$ and $r_1(f)$ be residuals under the null ($\beta=0$) and full (MLE) models. Define
\begin{align}
G_k(f) &= \sum_{t=1}^{T} \bigl|r_{k,t}(f)\bigr|^{2}, \label{eq:Gk}\\
Q_k(f) &= \sum_{t=1}^{T} r_{k,t}(f)_t^{\,2}, \label{eq:Ck}\\
P_k(f) &= \frac{G_k(f)^2 - \big|Q_k(f)\big|^{2}}{G_k(f)}, \qquad k\in\{0,1\}. \label{eq:Pk}
\end{align}

The per-band deviance is
\begin{equation}
\label{eq:glm-deviance}
D(f) = T \left[
  \log\!\frac{G_0(f)}{G_1(f)} \;+\;
  \log\!\frac{P_0(f)+\varepsilon}{P_1(f)+\varepsilon}
\right].
\end{equation}

which is asymptotically $\chi^2$-distributed \cite{wilks1938large} with $\mathrm{df}=2$ for one complex predictor (real and imaginary parts). The $p$-value is
\[
p_{\mathrm{GLM}}(f)=1-F_{\chi^2_2}\!\big(D(f)\big).
\]
Here $T$ is the number of DBT samples in the band, and $\varepsilon$ is a small constant (MATLAB \texttt{eps} in our implementation) matching the code’s stability term.

\textbf{Nonparametric (surrogates).}
Let $A_X(t,f)$ and $A_Y(t,f)$ be the complex DBT band representations (``blrep'') of the driver and observation with DBT sample index $t=1,\ldots,T$ and band $f$.
For each permutation $p=1,\dots,n_{\mathrm{perm}}$ (we use $n_{\mathrm{perm}}=2000$ throughout), we generate spectrum–preserving surrogates of $A_X$ while keeping $A_Y$ fixed:

\emph{(i) Circular shift:} draw a lag $\ell_p\sim\mathrm{Unif}\{0,\ldots,T-1\}$ and set
\[
\widetilde{A}_X^{(p)}(t,f)=A_X\!\big( (t+\ell_p)\bmod T,\; f\big).
\]

\emph{(ii) Phase randomization:} multiply by independent phases,
\[
\widetilde{A}_X^{(p)}(t,f)=A_X(t,f)\,e^{i\theta_{t,f}^{(p)}},\qquad \theta_{t,f}^{(p)}\sim \mathrm{Unif}[0,2\pi).
\]

For each surrogate, the coherence amplitude is
\[
C^{(p)}(f) = \frac{\big|\sum_t \widetilde{A}_X^{(p)}(t,f)\,A_Y^*(t,f)\big|}
{\sqrt{\big(\sum_t|\widetilde{A}_X^{(p)}(t,f)|^2\big)\big(\sum_t|A_Y(t,f)|^2\big)}}.
\]

Denote the observed value $C(f)$ and the surrogate values $C_{\text{circ}}^{(p)}(f)$, $C_{\text{phase}}^{(p)}(f)$. One-sided (upper-tail) empirical $p$-values are
\begin{align}
p_{\text{circ}}(f)  &= \tfrac{1}{n_{\mathrm{perm}}}\sum_{p=1}^{n_{\mathrm{perm}}} \mathbb{I}\!\big(C(f)\le C_{\text{circ}}^{(p)}(f)\big), \label{eq:pcirc}\\
p_{\text{phase}}(f) &= \tfrac{1}{n_{\mathrm{perm}}}\sum_{p=1}^{n_{\mathrm{perm}}} \mathbb{I}\!\big(C(f)\le C_{\text{phase}}^{(p)}(f)\big). \label{eq:pphase}
\end{align}
This induces a resolution floor $p_{\min}=1/(n_{\mathrm{perm}}{+}1)$. Ties are rare at this resolution and were handled implicitly by the empirical average.%
\footnote{Within a narrow analytic band, a circular time shift is equivalent to a constant phase rotation; both surrogates preserve the marginal spectrum while disrupting time-locked cross-dependence.}

%\subsection{Power Analysis}
%To estimate detection power, we controlled false discoveries across \emph{frequencies} within each simulation.For each realization $i$ at a given true coherence $C_{\mathrm{true}}$ injected at the breathing frequency $f_{\mathrm{br}}$, DBT produced coherence (and corresponding $p$-values) across center frequencies $\{f_k\}_{k=1}^{K}$. For each method $m\in\{\mathrm{GLM}, \mathrm{circ}, \mathrm{phase}\}$, we applied Benjamini--Hochberg FDR at level $\alpha=0.05$ to the set $\{p_i^{m}(f_k)\}_{k=1}^{K}$ and declared a detection when the breathing bin survived FDR:
%\[\mathbbm{1}^{m}_i = \mathbb{I}\!\left\{\, p^{m}_i(f_{\mathrm{br}})\ \text{is rejected under BH-FDR at } \alpha=0.05 \,\right\}.\]
%Power curves were obtained by sweeping $C_{\mathrm{true}}$ over 9{,}999 levels, binning $\approx 250$ simulations per coherence bin (bin width $\approx 0.025$), and summarizing by the thresholds at 50\%, 80\%, and 90\% detection (C50/C80/C90). We also report the empirical $\mathrm{SNR}_{\mathrm{dB}} = 10\log_{10}\!\big(\mathrm{var}(x)/\mathrm{var}(n)\big)$ at 80\% power.

\subsection{Power Analysis}
To estimate detection power, we defined significance at $\alpha=0.05$ using raw $p$-values. 
For each realization $i$ at a given true coherence $C_{\mathrm{true}}$ injected at the breathing frequency $f_{\mathrm{br}}$, we evaluated detection indicators
\[
\mathbbm{1}^{m}_i = \mathbb{I}\!\left\{\, p^{m}_i(f_{\mathrm{br}}) < 0.05 \,\right\}, 
\qquad m\in\{\mathrm{GLM}, \mathrm{circ}, \mathrm{phase}\}.
\]
Power curves were obtained by sweeping $C_{\mathrm{true}}$ over 9{,}999 levels, binning $\approx 250$ simulations per coherence bin (bin width $\approx 0.025$), and summarizing by the thresholds at 50\%, 80\%, and 90\% detection (C50/C80/C90). 
We also report the empirical $\mathrm{SNR}_{\mathrm{dB}} = 10\log_{10}\!\big(\mathrm{var}(x)/\mathrm{var}(n)\big)$ at 80\% power.

\subsection{ROC Analysis}
To quantify discriminability, ROC curves were constructed by treating the breathing frequency $f_{\mathrm{br}}$ as positives and a nearby control frequency (1\,Hz) as negatives.  
The area under the curve (AUC) was computed from the empirical hit/false-alarm rates.

\subsection{Runtime benchmarking}
We benchmarked three significance tests—GLM (parametric), circular shift (surrogate), and phase randomization (surrogate)—under identical DBT inputs. To keep timings tractable for visualization, we used a subsampled target set with $N_{\text{coh}}=100$ equally spaced coherence levels (a slice of $A_Y$) and fixed the surrogate permutation count at $n_{\mathrm{perm}}=2000$. For the single-setting comparison at $n_{\mathrm{perm}}=2000$, each method was timed across $n=10$ repeats (after a warm-up run) using MATLAB’s \texttt{timeit}; we report median~$\pm$~standard deviation. Speedups are reported as the ratio of median runtimes relative to each surrogate method. We also profiled scaling with the number of surrogates by varying $n_{\mathrm{perm}}\in\{100,200,500,1000,2000,4000\}$ on the same subsampled dataset. DBT coefficients were precomputed once and reused across methods (no file I/O); the cost of surrogate generation was included in the timings.

All benchmarks were run in MATLAB R2023b (PCWIN64) on Windows~10 with an 11th Gen Intel\textsuperscript{\textregistered} Core\textsuperscript{TM}
i7--11850H CPU (8 cores / 16 threads, 2.50~GHz) and 32~GB RAM. All runs
executed on CPU only with \texttt{maxNumCompThreads=8}. The full dataset dimensions were $\mathrm{AX}=148\times 32$ and $\mathrm{AY}=148\times 32\times 9{,}999$ (coherence targets from $0.0001$ to $0.9999$ in steps of $10^{-4}$); for runtime experiments we used the subsampled target set $\mathrm{AY}_{\text{sub}}$ of size $148\times 32\times 100$ ($N_{\text{coh}}=100$) drawn from the same $\mathrm{AY}$ tensor.

%=====================================================================
\section{Results}\label{sec:results}

\subsection{Observed vs.\ True Coherence}
\begin{figure}[t]
  \centering
  \includegraphics[width=\linewidth]{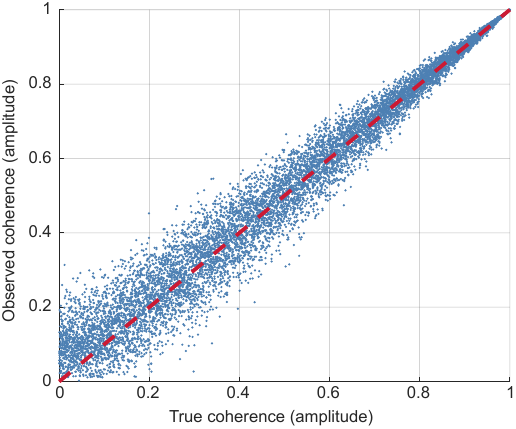}% True vs. measured scatter
  \caption{Observed (DBT) coherence amplitude at the frequency bin $f=0.3\,Hz$ versus true coherence across simulations. The dashed identity line highlights the close mapping from true to observed coherence under the broadband noise calibration.}
  \label{fig:cal}
\end{figure}

Figure~\ref{fig:cal} shows that DBT-estimated observed coherence at the breathing frequency increases monotonically with the injected (true) coherence across the full range, with modest dispersion around the identity due to finite-sample variability under broadband noise. 

%To illustrate what the tests “see,” Figures~\ref{fig:surr_low}–\ref{fig:surr_high} plot surrogate null histograms  ( $n_{\mathrm{perm}}{=}2000$ ) for three representative couplings: true $C_{\mathrm{true}}{=}0.18$ (low), $0.45$ (mid), and $0.85$ (high). The red vertical line marks the observed coherence statistic for that simulation. At $C_{\mathrm{true}}{=}0.18$, both GLM and surrogates are non-significant; at $C_{\mathrm{true}}{=}0.45$, GLM is strongly significant while the surrogate $p$ is not; and at $C_{\mathrm{true}}{=}0.85$ both tests are decisively significant. These snapshots foreshadow the group results (power vs.\ coupling, significance vs.\ SNR, and $p$-value agreement) reported next.

\subsection{Detection power versus coupling}
% \begin{figure}[t]
%     \centering
%     \includegraphics[width=0.95\columnwidth]{figures/FigPower.pdf}
%     \caption{Detection power (FDR-corrected $p<0.05$) as a function of true coherence amplitude. GLM (orange) and phase randomization (green) achieve higher power at lower coherence values compared to circular shift surrogates (blue). Thresholds for 50\%, 80\%, and 90\% detection (C50/C80/C90) are summarized in Table~\ref{tab:power}.}
%     \label{fig:power}
% \end{figure}

\begin{figure}[t]
    \centering
    \includegraphics[width=0.95\columnwidth]{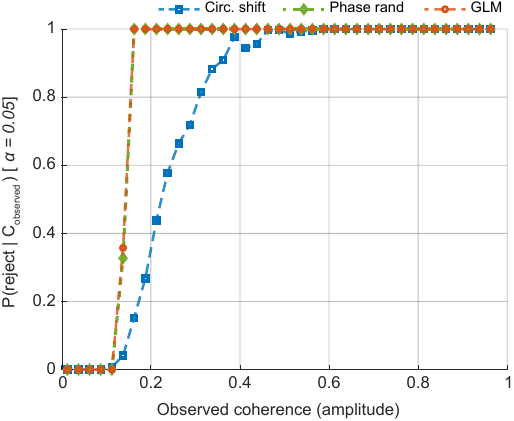}
    \caption{Detection power (raw $p<0.05$) as a function of \emph{observed} coherence amplitude at the breathing bin. GLM (orange) and phase randomization (green) achieve higher power at lower observed coherence than circular-shift surrogates (blue). Thresholds for 50\%, 80\%, and 90\% detection (C50/C80/C90) are summarized in Table~\ref{tab:power}.}
    \label{fig:power}
\end{figure}
Figure~\ref{fig:power} shows empirical detection rate versus \emph{observed} coherence $C_{\mathrm{obs}}$ at the breathing bin using raw $p<0.05$. Curves were obtained by binning $\approx250$ targets per $C_{\mathrm{obs}}$ bin (width $\approx0.025$) and averaging within bins; C50/C80/C90 were read off as first threshold crossings. Both GLM and phase randomization reached a given power level at substantially lower observed coherence than circular shift. For example, 80\% power required $C_{\mathrm{obs}}{=}0.16$ for GLM and phase randomization versus $0.31$ for circular shift. In SNR terms, this corresponds to $\text{SNR}@80\% = -45.8$\,dB (GLM), $-45.8$\,dB (phase), and $-37.4$\,dB (circular shift), i.e., GLM achieves the same power under $\sim$8.4\,dB lower SNR than circular shift, with phase randomization matching GLM ($\Delta=0.0$\,dB).

%Figure~\ref{fig:power} shows empirical detection rate versus true coherence ($C_{\mathrm{true}}$), sweeping from 0.001 to 0.9999 in 0.0001 increments ($9{,}999$ targets). A detection was counted when the FDR-corrected $p$ at the breathing frequency ($f_{\mathrm{br}}$) was $<0.05$. Curves were obtained by binning $\approx250$ targets per $C$ bin (bin width $0.025$) and averaging within bins; C50/C80/C90 were read off as the first threshold crossings. Both GLM and phase randomization reached a given power level at substantially lower coherence than circular shift. For example, 80\% power required $C{=}0.25$ (GLM) and $0.23$ (phase) versus $0.49$ (circular shift). In SNR terms, this corresponds to $\text{SNR}@80\% = -38.5$\,dB (GLM), $-39.3$\,dB (phase), and $-31.8$\,dB (circular shift), i.e., GLM achieves the same power under $\sim$6.8\,dB lower SNR than circular shift, with phase randomization slightly outperforming GLM ($\Delta=-0.8$\,dB).

% \begin{table}[t]
% \centering
% \caption{Power thresholds and SNR at 80\% power using \textbf{FDR-corrected} $p{<}0.05$ 
% (lower is better for $C$; more negative is better for SNR).}
% \label{tab:power}
% \begin{tabular}{lccc}
% \toprule
%  & Circ.\ shift & Phase rand & GLM \\
% \midrule
% C50               & 0.39 & 0.17 & 0.17 \\
% C80               & 0.49 & 0.23 & 0.25 \\
% C90               & 0.55 & 0.27 & 0.27 \\
% SNR@80\% (dB)     & $-31.8$ & $-39.3$ & $-38.5$ \\
% \bottomrule
% \end{tabular}
% \end{table}

\begin{table}[t]
\centering
\caption{Power thresholds and SNR at 80\% power using \textbf{raw} $p{<}0.05$ (lower is better for $C$; more negative is better for SNR).}
\label{tab:power}
\begin{tabular}{lccc}
\toprule
 & Circ.\ shift & Phase rand & GLM \\
\midrule
C50               & 0.24 & 0.16 & 0.16 \\
C80               & 0.31 & 0.16 & 0.16 \\
C90               & 0.36 & 0.16 & 0.16 \\
SNR@80\% (dB)     & $-37.4$ & $-45.8$ & $-45.8$ \\
\bottomrule
\end{tabular}
\end{table}

Thus, GLM and phase randomization provide markedly higher sensitivity than circular shift—requiring $\sim$0.20–0.25 lower coherence and $\sim$6–7\,dB lower SNR to achieve 80\% detection power—while showing comparable performance near 50\% power.

\begin{figure}[t]
    \centering
    \includegraphics[width=0.8\columnwidth]{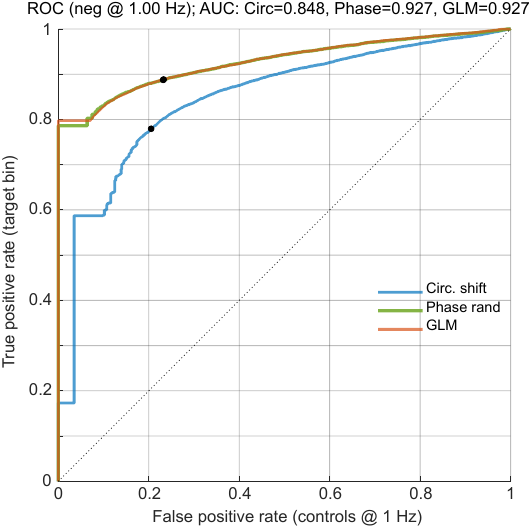}
    \caption{Receiver operating characteristic (ROC) analysis of detection methods. ROC curves were constructed by treating coherence at the breathing frequency as positives and coherence at a nearby control frequency (1.0\,Hz) as negatives. GLM (orange) and phase randomization (green) achieved superior discriminability (AUC $= 0.927$ for both) compared to circular shift surrogates (blue; AUC $= 0.848$). The dotted diagonal indicates chance-level performance.}
    \label{fig:roc}
\end{figure}

To further quantify discriminability, we compared coherence estimates at the breathing frequency against a nearby control frequency (1.0\,Hz) using ROC analysis (Fig.~\ref{fig:roc}). 
GLM and phase-randomization methods both achieved high classification performance (AUC = 0.927), indicating strong separation of breathing-related coherence from the control frequency. By contrast, the circular-shift method performed less well (AUC = 0.848), reflecting greater overlap between positive and control bins. At the conventional threshold of $p=0.05$, GLM and phase randomization correctly identified approximately 75–80\% of true breathing bins while maintaining a false positive rate below 10\%, whereas the circular-shift method yielded lower sensitivity. These results demonstrate that GLM and phase randomization are more effective than circular shift for detecting respiratory–neural coherence at the single-frequency level.

% \subsection{Significance versus SNR}
% \begin{figure}[t]
%     \centering
%     \includegraphics[width=0.95\columnwidth]{Fig6.png}
%     \caption{Significance vs.\ SNR. Points show FDR-corrected $p$-values (plotted as $-\log_{10}p$) for surrogate test (blue) and the GLM (orange). The horizontal dashed line marks the decision threshold $p=0.05$ ($-\log_{10}p\approx1.30$). The dotted line marks the surrogate resolution floor $p_{\min}=1/(n_{\mathrm{perm}}{+}1)=1/2001\approx 0.0005$ ($-\log_{10}p\approx3.30$). GLM points that underflow to $p{=}0$ after FDR are plotted at the y-axis cap (here $-\log_{10}p = 4$). At 80\% power, GLM requires lower SNR (GLM: $-40.9$\,dB; Surrogate: $-38.9$\,dB; $\Delta=2.0$\,dB).}    
%     \label{fig:snr}
% \end{figure}
% Figure~\ref{fig:snr} relates broadband SNR ($10\log_{10}(\mathrm{var}(x)/\mathrm{var}(n))$) to $-\log_{10}p$ after FDR. GLM exhibits a sharper “knee’’ and larger $-\log_{10}p$ at a given SNR. Consistent with Fig.~\ref{fig:power}, GLM attains 80\% power at $\sim$2.0\,dB lower SNR (about $1.6\times$ lower on a linear scale).

\subsection{P-value Agreement and Permutation Floor}
\begin{figure}[t]
    \centering
    \includegraphics[width=0.9\columnwidth]{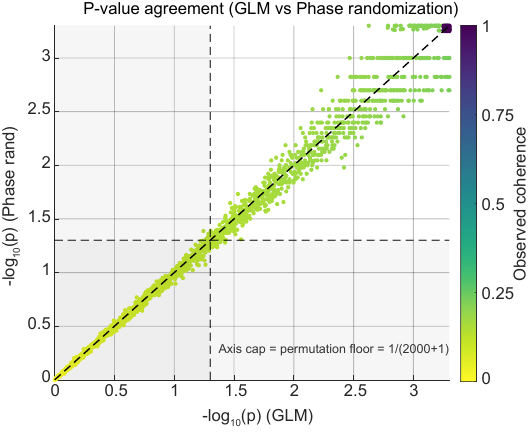}
    \caption{P-value agreement between GLM and the phase-randomization surrogate: $-\log_{10}p_{\mathrm{phase}}$ (y) vs.\ $-\log_{10}p_{\mathrm{GLM}}$ (x). Each point is one simulated case; color indicates observed coherence amplitude. The dashed diagonal is the identity line. Vertical/horizontal dashed lines mark $p=0.05$ ($-\log_{10}p\approx1.30$). Because permutation tests cannot yield $p<1/(n_{\mathrm{perm}}{+}1)$ ($n_{\mathrm{perm}}{=}2000\Rightarrow -\log_{10}p_{\min}\approx 3.30$), surrogate values saturate near 3.30. For visualization parity, GLM values exceeding this level were \emph{clipped} at 3.30, producing the dense cluster near the top-right; these correspond to high-coherence cases significant by both methods.}
    \label{fig:agree}
\end{figure}

Figure~\ref{fig:agree} compares the raw $p$ values of GLM with those of phase randomization (the strongest surrogate baseline). Most points fall below the identity line, showing GLM typically yields smaller (more significant) $p$ than phase randomization for the same coherence. Because permutation tests cannot report values below the resolution floor $p_{\min}=1/(n_{\mathrm{perm}}{+}1)$ ($n_{\mathrm{perm}}=2000 \Rightarrow -\log_{10}p_{\min}\approx 3.30$), surrogate $-\log_{10}p$ saturates for strong effects, whereas GLM continues to increase smoothly. For visualization parity, GLM values exceeding the surrogate floor were clipped at this display cap, producing the dense cluster near the top-right corner; these points correspond to high-coherence cases that are significant by both methods.

\subsection{Runtime Performance}
At $n_{\mathrm{perm}}=2000$ (subsampled $N_{\text{coh}}=100$), runtimes over 10 runs were
8.022~s $\pm$ 0.411~s for circular shift, 7.558~s $\pm$ 0.331~s for phase randomization, and 0.041~s $\pm$ 0.002~s for GLM (median~$\pm$~std). This corresponds to $\sim 198\times$ and $\sim 186.5\times$ speedups for GLM relativeto circular shift and phase randomization, respectively. Varying the number of surrogates showed near-linear scaling for both surrogate tests, while GLM remained essentially flat.

\begin{figure}[t]
    \centering
    \includegraphics[width=0.95\columnwidth]{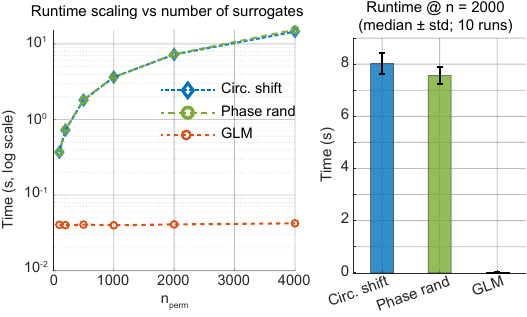}
    \caption{\textbf{Runtime comparison.}
    (\textbf{Left}) Runtime scaling versus number of surrogates ($n_{\mathrm{perm}}$; log scale) on the subsampled dataset ($N_{\text{coh}}=100$). Surrogate methods (circular shift, phase randomization) scale nearly linearly with $n_{\mathrm{perm}}$, whereas GLM is flat.  (\textbf{Right}) Runtimes at $n_{\mathrm{perm}}=2000$ over 10 runs (median~$\pm$~std): circular shift $8.022 \pm 0.411$\,s, phase randomization $7.558 \pm 0.331$\,s, GLM $0.041 \pm 0.002$\,s ($\sim198\times$ and $\sim186.5\times$ faster than the surrogates, respectively).}
    \label{fig:runtime}
\end{figure}

%==================================================================
%************************** DISCUSSION ****************************
%==================================================================
\section{Discussion}\label{sec:discussion}
We evaluated a generalized linear model (GLM) applied to complex DBT coefficients as a parametric alternative to surrogate-based significance tests for spectral coherence. GLM matched the sensitivity of phase randomization, the surrogate approach most often recommended for coherence, while outperforming circular-shift surrogates across coupling levels. GLM also produced smooth, continuous $p$-values and avoided the resolution floor imposed by finite permutations (e.g., $p_{\min}=1/(n_{\mathrm{perm}}+1)$).

We quantified computational efficiency under identical inputs. At $n_{\mathrm{perm}}=2000$ on the subsampled set ($N_{\text{coh}}=100$), runtimes over 10 runs (median~$\pm$~SD) were: circular shift $8.022\pm0.411$ s, phase randomization $7.558\pm0.331$, s and GLM $0.041\pm0.002$, s. 
This corresponds to $\sim$198$\times$ speedup relative to circular shift and $\sim$186.5$\times$ relative to phase randomization. Nonparametric tests on log-times confirmed these differences (Wilcoxon $p=0.00195$ for GLM vs.\ circular shift and GLM vs.\ phase; paired $t$-tests on log-times $p<10^{-6}$ for both). These results indicate that GLM delivers nearly two orders of magnitude faster inference without compromising detection performance, which is advantageous for multichannel EEG/iEEG, high sampling rates, or large frequency grids.

The limitations are: we simulated Gaussian broadband noise with fixed DBT parameters, so thresholds may vary under different spectra or time–bandwidth products; we used respiration as the only driver, and extensions to other signals such as cardiac or speech remain to be tested; the GLM assumes complex normal errors and approximate stationarity, which may not always hold; and runtime numbers came from CPU tests on a subsampled dataset, so absolute times will change with hardware and dataset size. Despite these constraints, the advantages of GLM—continuous inference, absence of permutation limits, and major computational savings—stayed consistent.

In general, GLM-based coherence testing offered a statistically valid and computationally efficient alternative to surrogate methods. It matched the sensitivity of phase randomization while eliminating resampling overhead, which makes it well suited for scalable analyses of brain–body and brain–signal interactions.

%==============================================================================================
\section{Conclusion}\label{sec:conclusion}
We introduced and validated a parametric framework for coherence significance testing that fits a generalized linear model (GLM) directly to complex DBT coefficients and evaluated it against spectrum-preserving surrogates under matched simulations with real respiration drivers. GLM matched the detection performance of the recommended phase randomization surrogate \cite{Faes2004} and outperformed the circular shift surrogates, while eliminating the permutation floor and yielding smooth, continuous $p$ values. Critically, GLM delivered more than two orders of magnitude speed-up relative to surrogate approaches, making it especially well suited for large-scale multichannel EEG and intracranial EEG at high sampling rates, where rapid mass-univariate inference across channels and frequencies is required.

\section*{Software Availability}
The code for reproducing the simulations, figures and GLM / coherence analyses is available at
\url{https://github.com/ckovach/ComplexGLM_paper}.

%\section*{References}
\bibliographystyle{IEEEtran}
\bibliography{references} % .bib file 

\end{document}